# Ultrafast dynamics of non-equilibrium electrons and strain generation under femtosecond laser irradiation of Nickel

George D.Tsibidis *

*Institute of Electronic Structure and Laser (IESL), Foundation for Research and Technology (FORTH), N. Plastira 100, Vassilika Vouton, 70013, Heraklion, Crete, Greece*

We present a theoretical study of the ultrafast electron dynamics in transition metals using ultrashort pulsed laser beams. The significant influence of the contribution of the dynamics of produced nonthermal electrons to electron thermalisation and electron-phonon interaction is thoroughly investigated within a range of values of the pulse duration (i.e. from 10fs to 2.3ps). The theoretical model elaborates firstly on possible transient changes of optical parameters during irradiation due to the presence of the non-equilibrium electrons and the modification of the electron distribution function in a material characterized by a variable electron density of states around the Fermi energy. The model correlates the role of nonthermal electrons, relaxation processes and induced stress-strain fields. Simulations are presented by choosing Nickel (Ni) as a test material to compute electron-phonon relaxation time due to its large electron-phonon coupling constant that expects to reveal more conclusively the influence of the non-equilibrium electrons. We demonstrate the consideration of the above factors leads to significant changes compared to the results the traditional Two Temperature Model (TTM) provides. The proposed model predicts a substantially (~33%) smaller damage threshold and a large increase of the stress (~20%, at early times) which firstly underlines the role of the nonthermal electron interactions and secondly enhances its importance with respect to the precise determination of laser specifications in material micromachining techniques.

## I. INTRODUCTION

Material processing with ultra-short pulsed lasers has received considerable attention over the past decades due to its important technological applications, in particular in industry and medicine [1-9]. These abundant applications require a thorough knowledge of the fundamentals of laser interaction with the target material for enhanced controllability of the resulting modification of the target relief. Physical mechanisms that lead to surface modification have been explored both theoretically and experimentally [10-23].

It is well-known that after irradiation, the initial electron population is highly nonthermal. Experimental observations related to the nonequillibrium dynamics of electron systems in metals shows nonthermalised electron distribution leads to a faster electron-phonon relaxation [24, 25]. Hence, to discriminate nonthermal electron-generated electron-phonon relaxation, it is necessary to investigate thoroughly the dynamics of nonthermal electron gas in these conditions.

On the other hand, and regarding modelling of the laser-matter interaction, it is known that the traditional Two Temperature Model (TTM) [26] assumes a rapidly (instantaneous) thermalisation of the electronic distribution. Therefore, TTM yields an overestimation of the electronic temperature which has been also confirmed by pump-probe experiments [27]. To overcome the limitations of the TTM, analysis based on the Boltzmann's transport equations [28] or revised versions of the TTM [27, 29-31] have been proposed. Those works presented the necessity for the inclusion of the initial nonthermal (NTH) electron population. More specifically, Sun at al. used simultaneously a three coupled equation-based model and Boltzmann's transport equation where heating of the electron gas by the initial NTH electrons and the lattice are included [31]. On the other hand, Lisowksi et al. proposed an improved version of the TTM by presenting a three temperature model where a temperature for the NTH electrons was introduced [27]. Although these approaches successfully predicted results from pump-probe experiments for noble metals [31] or Ruthenium [27], they were characterised by the necessity of including fitting parameters.

By contrast, in a recent approach presented by Carpene, an extension of the TTM was introduced by incorporating the electron thermalisation dynamics into the source term, a potentially direct energy transfer from NTH electrons to the lattice and the consideration of a very-low-density NTH electron distribution [29]. The approach allows a consistent calculation of the resulting (thermalized) electron temperature assuming the contribution of both the thermalized (TH) and NTH electron distributions. The incorporation of the influence of both TH and NTH electron dynamics, also, overcomes the weakness of the classical TTM related to the nonequilibrium state of the NTH, and thereby, inability to define an "electron temperature" in the early stages. Validation of the model through pump-probe experiments illustrated that it provides accurate description of ultrafast dynamics after irradiation of noble metals with ultrashort laser pulses [32-36].

Nevertheless, there are still some open questions regarding the model presented by Carpene. It was assumed that the electron density of states (DOS) is constant around the Fermi energy [29] which constitutes a sufficient approximation for noble metals (i.e. Au, Cu, Ag), however, for transient metals such as Nickel (Ni) or Titanium (Ti) it is a rather crude approximation that expects to lead to inaccurate results (see Fig.1) [37]; hence, a more rigorous approach is required to determine the combined influence of the NTH electrons and DOS. On the other hand, experimental observations related to the nonequillibrium dynamics of electron systems in metals shows NTH electron distribution leads to a faster electron-phonon relaxation [24, 25]. It is important, thus, to elaborate on how metals characterised by a large electron-phonon influence relaxation processes.

The elucidation of the aforementioned features is of paramount importance not only to understand further the underlying physical mechanisms of laser-matter interactions and ultrafast electron dynamics but also to associate the resulting thermal effects with the surface response which can be used to process systematically the material. Therefore, there is a growing interest to reveal the physics of the underlying processes from both a fundamental and application point of view.

To proceed with the influence of electronic excitation on the morphological changes, one aspect that has yet to be explored is the correlation of the pulse width, energy deposition, optical properties changes during the pulse duration and structural changes. In principle, morphological surface changes at low excitation levels are strongly related to stress generation as well as whether lattice temperatures induce large stresses (i.e. that exceeds the yield stress) [38]. It is known that although NTH interactions with TH electrons and the lattice yield remarkably smaller maximum electron temperature they induce smaller maximum lattice temperature variations due to the large lattice heat capacity [29]. Nevertheless, it is important to investigate whether a material with a large electron-phonon coupling constant and a complex DOS around the Fermi energy behaves differently and thereby, significant lattice temperature changes occur which, in turn, is also reflected on an enhanced mechanical response.

To this end, we present an extension of the model proposed by Carpene which comprises: (i) the influence of the DOS around the Fermi energy, (ii) a component that corresponds to the temporal evolution of the optical parameters during the pulse duration, (iii) a thermomechanical component that describes the mechanical response of the material due to material heating. In order to highlight the significance of the nonthermal electrons for materials with large and temperature dependent electron-lattice coupling constant, $G_{eL}$, Ni is used.

For the sake of simplicity, the investigation has been focused on single shot while a similar approach could be pursued in case of repetitive irradiation. Low laser fluences have been used primarily in this work to distinguish the role of nonthermal electron contribution and electron-phonon relaxation processes; this is due to the fact that at larger energies more complex effects such as ablation or melting occur. In that case, hydrodynamical models or atomistic simulations are required to be incorporated into a multiscale theoretical framework which could hinder the significance of the aforementioned factors while possible major morphological changes might also be attributed to other effects [10, 39, 40]. Nevertheless, prediction for the electronic and lattice temperatures at larger fluences were also performed to estimate the energy fluence at which damage occurs (i.e. when lattice temperature exceeds the melting point of the material).

In the following section, we present the theoretical framework used to describe the physical mechanism and the components of the revised TTM (rTTM). Section III explains the numerical algorithm and the adaptation of the model to Ni. A systematic analysis of the results and the role of the nonthermal electron contribution are presented in Section IV while the energy exchange between electrons and lattice for small excitations is also investigated. We determine the thermalisation time of the laser-excited electron gas as a function of the pulse duration. The resulting morphological changes are correlated with the pulse duration through the computation of strain fields and displacements. A parametric study is followed in which the role of pulse duration variation in both thermal (evolution of electron and lattice temperatures) and mechanical response of the material (stress/strain propagation) is investigated. All theoretical results are tested against the traditional TTM to highlight the discrepancies. Furthermore, simulations yield the magnitude of change of the optical characteristics within the laser heating time that potentially affects laser energy absorption. Fluence dependence of the thermomechanical response and determination of ablation thresholds are also explored. Concluding remarks follow in Section V.

## II. THEORETICAL MODEL

### A. Laser Beam Profile and Non-equilibrium Electrons

The laser pulse at time $t'$ is described by an energy flux (in a three dimensional space characterised by the Cartesian coordinates $x,y,z$), $I(t,x,y,z)$ provided by the following expressions

$$-\frac{\partial I(t',x,y,z)}{\partial z} = \alpha(t',x,y,z)I(t',x,y,z) = W(t',x,y,z) \quad (3)$$

$$W(t',x,y,z=0) = (1-R(t',x,y,z=0)) \frac{2\alpha\sqrt{\log(2)}}{\sqrt{\pi}\tau_p} E_p \times$$
$$\exp\left(-4\log(2)\left(\frac{t'-3\tau_p}{\tau_p}\right)^2\right) \times \exp\left(-\frac{x^2+y^2}{(R_0)^2}\right) \quad (4)$$

where $W(t',x,y,z)$ corresponds to the absorbed laser power density at time $t'$, $\alpha$ is the absorption coefficient, $\tau_p$ is the pulse duration, $E_p$ is the fluence and $R_0$ stands for the irradiation spot radius. For the sake of simplicity, it is assumed that the ballistic length of electrons for Ni is small and therefore it is neglected [41]. Based on a previous work by Carpene [29], it is assumed that within an infinitesimal time $\Delta t'$ laser photons will interact with electrons lying in occupied states below Fermi energy $\varepsilon_F$ causing their excitation to previously unoccupied states above $\varepsilon_F$. As result, an infinitesimal nonthermal change $\Delta f$ to the (Fermi-Dirac) electronic distribution $f(\varepsilon)$ will be produced which has the form [29, 31]

$$\Delta f(\varepsilon,t',x,y,z) = \delta_0(t',x,y,z)\big(f(\varepsilon-h\nu) \times [1-f(\varepsilon)] - f(\varepsilon)[1-f(\varepsilon+h\nu)]\big) \quad (5)$$

where $f(\varepsilon) = \{1+\exp[(\varepsilon-\mu_f)/(k_B T_e)]\}^{-1}$ is the Fermi-Dirac distribution for electrons with energies $\varepsilon$ and temperature $T_e$, $h\nu$ is the laser photon energy (~1.55eV for laser beam wavelength $\lambda_L$=800nm), $\mu_f$ is the chemical potential that is equal to $\varepsilon_F$ at $T_e$=0K and $k_B$ is the Boltzmann constant. We note that comparing the underlying physics with the proposed processes in the case of noble metals [29], some significant changes have to be performed to describe nonthermal electron distribution and their influence in excitation, thermalisation and electron-phonon relaxation processes for transient materials (i.e. Ni or Ti). More specifically, it is evident (Fig.1) that at fluences large enough to produce very energetic electrons, the variation of the Fermi-Dirac distribution cannot be described by a step-like function for energies in the range [$\varepsilon_F$-$h\nu$, $\varepsilon_F$+$h\nu$] [29]. It is noted that results in Fig.1 are normalised to 1 (through the parameter $\delta_0$); it is evident that at larger temperatures a different shape of $\Delta f$ contains a substantially large tail that becomes even bigger with increasing $T_e$. According to Fig.1, the non-zero change of the Fermi-Dirac distribution in a larger range of energies (characteristic for larger electron temperatures) requires consideration of the complete expression of the nonthermal electron distribution to estimate the correlation of the energy density associated with the nonthermal electrons and the absorbed laser power density. The size of $\Delta f$ is computed by the following expression

$$\int_{\varepsilon_F-h\nu}^{\varepsilon_F+h\nu} \Delta f(\varepsilon,t',x,y,z)N(\varepsilon)\varepsilon\, d\varepsilon = W(t',x,y,z)\, dt' \quad (6)$$

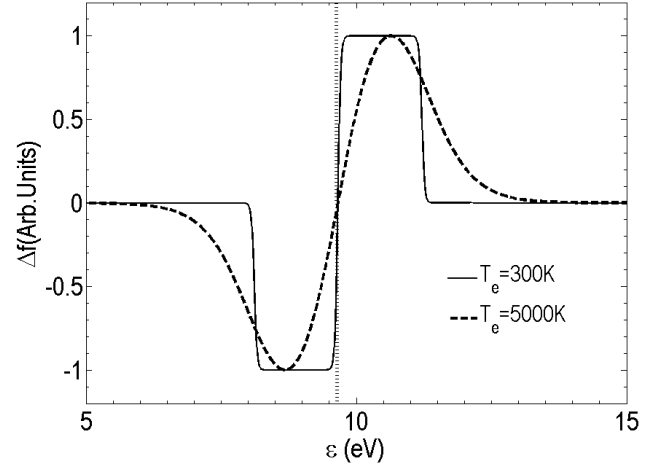

FIG. 1. Change of the Fermi-Dirac distribution function due to nonthermal electrons for $T_e$=300K (*solid* line) and $T_e$=5000K (*dashed* line). The *dotted* vertical line indicates the position of energy Fermi $\varepsilon_F$.

which relates the absorbed laser power density within the interval $\Delta t'$ with the energy density of the NTH electrons for energies in [$\varepsilon_F$-$h\nu$, $\varepsilon_F$+$h\nu$]. We note that the DOS $N(\varepsilon)$ is provided in Fig.2 based on simulations performed in Ref. [37]. Then, the analytical expression for $\delta_0$ that results from Eqs.(5,6) is the following

$$\delta_0(t',x,y,z) = \frac{W(t',x,y,z)\, dt'}{\int_{\varepsilon_F-h\nu}^{\varepsilon_F+h\nu} \big(f(\varepsilon-h\nu)[1-f(\varepsilon)] - f(\varepsilon)[1-f(\varepsilon+h\nu)]\big)N(\varepsilon)\varepsilon\, d\varepsilon} \quad (7)$$

It is evident that while it might be assumed that a constant DOS can be used for energies in the range [$\varepsilon_F$-$h\nu$, $\varepsilon_F$+$h\nu$] for noble metals [37], it is rather a crude approximation for transient metals such as Ni (Fig.2). Due to the lack of a simple analytical expression for $N(\varepsilon)$ for energies in that range, the computation of $\Delta f$ is derived through a numerical solution of Eq.(7). Fig.2 also indicates that the *d*-band electrons can easily excited even at low electron temperatures which is determined by both electron-electron and electron-phonon interactions.

On the other hand, the initial nonthermal electron distribution relaxes at a rate which is determined both from an electron-electron and electron-phonon interaction provided by the following expression [29]

$$\Delta f_{NT}(\varepsilon,t-t',x,y,z) = \exp\left(-\frac{t-t'}{\tau_{ee}} - \frac{t-t'}{\tau_{ep}}\right)\Delta f(\varepsilon,t',x,y,z) \quad (8)$$

where $\tau_{ee} = \frac{128}{\sqrt{3}\pi^2 \omega_p}\left(\frac{\varepsilon_F}{\varepsilon - \varepsilon_F}\right)^2 \equiv \tau_0 \left(\frac{\varepsilon_F}{\varepsilon - \varepsilon_F}\right)^2$ and $\omega_p$ is the plasma frequency (=15.92eV for Ni [42]) and $\varepsilon_F \simeq 9.7$eV for Ni [43]. By contrast, the electron-phonon collision is provided by the expression $\tau_{ep} = \frac{h\nu}{k_B \Theta_D}\tau_f$, where $\tau_f$ ($\simeq 2.25$fs for Ni [43]) stands for the time between two subsequent collisions with the lattice of nonthermal quasiparticles of energy $\varepsilon-\varepsilon_F \simeq h\nu$ while $\Theta_D$ ($\Theta_D = 450$K, for Ni [44]) is the Debye temperature.

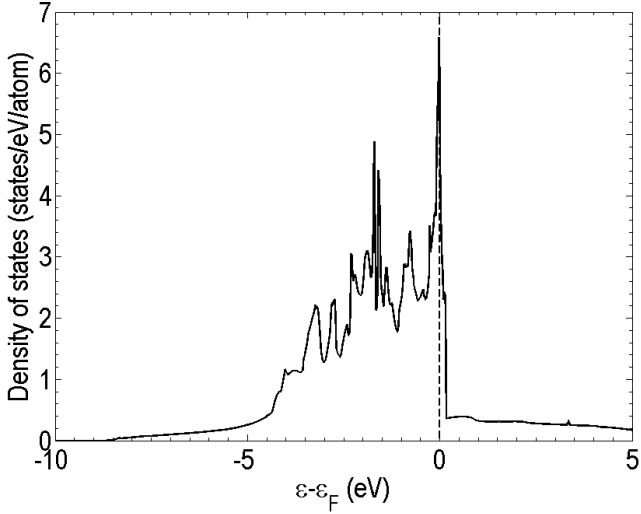

FIG. 2. The electron DOS of Ni (data calculated in Ref. [37]). The *dashed* vertical line indicates the position of the Fermi energy.

The rate of the energy density $u$ that is transferred from the nonthermal electron distribution to both the thermal electrons and the lattice is described by the following expression [29]

$$\frac{\partial u(t,t')}{\partial t} = \frac{\partial}{\partial t}\int_{\varepsilon_F - h\nu}^{\varepsilon_F + h\nu} \Delta f_{NT}(\varepsilon, t-t', x, y, z) N(\varepsilon)\varepsilon d\varepsilon$$
$$= \int_{\varepsilon_F - h\nu}^{\varepsilon_F + h\nu} \frac{\partial}{\partial t}\left(\Delta f_{NT}(\varepsilon, t-t', x, y, z)\right) N(\varepsilon)\varepsilon d\varepsilon$$
$$= -\int_{\varepsilon_F - h\nu}^{\varepsilon_F + h\nu} \left(\frac{1}{\tau_0}\left(\frac{\varepsilon - \varepsilon_F}{\varepsilon_F}\right)^2 + \frac{1}{\tau_{ep}}\right)\Delta f(\varepsilon, t', x, y, z) \times \quad (9)$$
$$\exp\left(-\frac{t-t'}{\tau_0}\left(\frac{\varepsilon - \varepsilon_F}{\varepsilon_F}\right)^2 - \frac{t-t'}{\tau_{ep}}\right) N(\varepsilon)\varepsilon d\varepsilon$$
$$\equiv \frac{\partial u_{ee}(t,t')}{\partial t} + \frac{\partial u_{eL}(t,t')}{\partial t}$$

where

$$\frac{\partial u_{ee}(t,t')}{\partial t} = -\frac{1}{\tau_0}\int_{\varepsilon_F - h\nu}^{\varepsilon_F + h\nu}\left(\frac{\varepsilon - \varepsilon_F}{\varepsilon_F}\right)^2 \Delta f(\varepsilon, t', x, y, z) \times$$
$$\exp\left(-\frac{t-t'}{\tau_0}\left(\frac{\varepsilon - \varepsilon_F}{\varepsilon_F}\right)^2 - \frac{t-t'}{\tau_{ep}}\right) N(\varepsilon)\varepsilon d\varepsilon \quad (10)$$

and

$$\frac{\partial u_{eL}(t,t')}{\partial t} = -\left(\frac{1}{\tau_{ep}}\right)\int_{\varepsilon_F - h\nu}^{\varepsilon_F + h\nu} \Delta f(\varepsilon, t', x, y, z) \times$$
$$\exp\left(-\frac{t-t'}{\tau_0}\left(\frac{\varepsilon - \varepsilon_F}{\varepsilon_F}\right)^2 - \frac{t-t'}{\tau_{ep}}\right) N(\varepsilon)\varepsilon d\varepsilon \quad (11)$$

The quantities $u_{ee}$ and $u_{eL}$ correspond to the rate of energy density exchanged between the NTH electrons and (i) the TH electrons and (ii) the lattice through electron-electron and electron-phonon scattering, respectively. Time $t$ corresponds to the timepoint at which the NTH distribution relaxes to $\Delta f$ at $t = t'$ as seen in Eq.8. The corresponding total energy relaxation (i.e. at time $t$) is derived by adding the infinitesimal contributions of excitation for time intervals $dt'$ at all times $t'<t$. As a result, the following expressions are obtained by integrating over $t'$,

$$\frac{\partial U_{ee}(t)}{\partial t} = \int_0^t \frac{\partial u_{ee}(t,t')}{\partial t} dt'$$
$$\frac{\partial U_{eL}(t)}{\partial t} = \int_0^t \frac{\partial u_{eL}(t,t')}{\partial t} dt' \quad (12)$$

### B. Dielectric constant

It is evident that the transient variation of the dielectric constant through the temperature dependence of the electron relaxation time will potentially produce a change of the optical properties of the material during irradiation. This variation cannot be ignored when the pulse duration is very small. A similar behaviour has been noted in noble metals (i.e. Au and Cu) where a static consideration of the optical properties led to an incorrect estimation of the energy deposition which was reflected from an underestimation of the thermal response of the material and the damage threshold [45, 46]. Therefore, a more complete approach should be based on a rigorous approach of considering a dynamic change of the optical properties.

To take into account temperature dependence of the optical characteristics and incorporate both interband and intraband transitions, the dielectric constant of Ni is modelled by means of an extended Lorentz-Drude model with four Lorentzian terms based on the analysis of Rakic *et* al. [42]

$$\varepsilon(\omega_L) = 1 - \frac{f_0 \omega_p^2}{\omega_L^2 - i\Gamma_0 \omega_L} + \sum_{j=1}^{k=4} \frac{f_j \omega_p^2}{\omega_j^2 - \omega_L^2 + i\omega_L \Gamma_j} \quad (13)$$

where $\omega_L$ is the laser frequency ($=2.3562 \times 10^{15}$ rad/s for 800nm that corresponds to photon energy equal to 1.55eV), and $\sqrt{f_0}\omega_p$ is the plasma frequency associated with oscillator strength $f_0$ and damping constant $\Gamma_0$. The interband part of the dielectric constant (third term in Eq.13) assumes four oscillators with frequency $\omega_j$, strength $f_j$, and lifetime $1/\Gamma_j$. Values for the aforementioned parameters are given in Ref.[42]. The damping constant $\Gamma_0$ is the reciprocal of the electron relaxation time, $\tau_e$, which is given by $\tau_e = 1/(B_L T_L + A_e(T_e)^2)$, where $T_e$, $T_L$ are the electron and lattice temperatures, respectively. Values of the coefficients $A_e$, $B_L$ are $0.59 \times 10^7$ (s$^{-1}$K$^{-2}$) and $1.4 \times 10^{11}$ (s$^{-1}$K$^{-1}$), respectively [47, 48]. The dynamic character of the optical parameters (i.e. refractive index $n$, extinction coefficient $k$, absorption coefficient $\alpha$, and reflectivity $R$) can be easily computed through the real and imaginary part of the dielectric constant $\varepsilon_1$ and $\varepsilon_2$, respectively [49]

$$\varepsilon(\omega_L, x, y, z, t) = \varepsilon_1(x,y,z,t) + i\varepsilon_2(x,y,z,t)$$

$$n = \sqrt{\frac{\varepsilon_1(x,y,z,t) + \sqrt{(\varepsilon_1(x,y,z,t))^2 + (\varepsilon_2(x,y,z,t))^2}}{2}}$$

$$k = \sqrt{\frac{-\varepsilon_1(x,y,z,t) + \sqrt{(\varepsilon_1(x,y,z,t))^2 + (\varepsilon_2(x,y,z,t))^2}}{2}} \quad (14)$$

$$\alpha(x,y,z,t) = \frac{2\omega_L k}{c}$$

$$R(x,y,z=0,t) = \frac{(1-n)^2 + k^2}{(1+n)^2 + k^2}$$

### C. Elasticity Equations

The mechanical response of the material is described by the differential equations of dynamic elasticity which correlate the stress and strain generation and the induced displacement as a result of the thermal expansion and the lattice temperature variation

$$\rho_L \frac{\partial^2 V_i}{\partial t^2} = \sum_{j=1}^{3} \frac{\partial \sigma_{ji}}{\partial x^j}$$

$$\sigma_{ij} = 2\mu\varepsilon_{ij} + \lambda \sum_{k=1}^{3} \varepsilon_{kk}\delta_{ij} - \delta_{ij}(3\lambda + 2\mu)\alpha'(T_L - T_0) \quad (15)$$

$$\varepsilon_{ij} = 1/2 \left( \frac{\partial V_i}{\partial x^j} + \frac{\partial V_j}{\partial x^i} \right)$$

where $V_i$ correspond to the displacements along the $x$ ($i=1$), $y$ ($i=2$), $z$ ($i=3$) direction, while $\sigma_{ij}$ and $\varepsilon_{ij}$ stand for the stresses and strains, respectively [50]. On the other hand, the Lame's coefficients $\lambda$ and $\mu$ (for Ni, $\lambda=1.25 \times 10^7$ and $\mu=7.63 \times 10^7$), respectively, $\alpha'$ stands for the thermal expansion of Ni and $\rho_L$ is the density of the material. The Lame's coefficients $\lambda$ and $\mu$ are related to the Poisson's ration ($\nu$) and Young's modulus ($E$) through the relations $\nu = \lambda/(2(\lambda+\mu))$ and $E = \mu(2\mu+3\lambda)/(\lambda+\mu)$.

### D. Generalised Energy Balance Equations

To describe the influence of the ultrafast electron dynamics the relaxation procedure, and the thermomechanical response of the material, a revised version of the TTM is used that includes the early and transient interaction of the non-thermal electron distribution with the electron and lattice baths [29, 30]. Hence, the following set of equations is employed to investigate the spatio-temporal distribution of the produced thermalized electron ($T_e$) and lattice ($T_L$) temperatures of the assembly

$$C_e \frac{\partial T_e}{\partial t} = \vec{\nabla} \cdot \left( k_e \vec{\nabla} T_e \right) - G_{eL}(T_e - T_L) + \frac{\partial U_{ee}}{\partial t}$$

$$C_L \frac{\partial T_L}{\partial t} = G_{eL}(T_e - T_L) - (3\lambda + 2\mu)\alpha' T_L \sum_{j=1}^{3} \dot{\varepsilon}_{jj} + \frac{\partial U_{eL}}{\partial t} \quad (16)$$

where the subscripts $e$ and $L$ are associated with electrons and lattice, respectively, $k_e$ ($=k_{e0}BT_e/(A_e(T_e)^2+B_L T_L)$) is the thermal conductivity of the electrons, $C_e$ and $C_L$ are the heat capacity of electrons and lattice, respectively, and $G_{eL}$ is the electron-phonon coupling factor. The parameters $C_e$ and $G_{eL}$ are taken from Ref. [37] (i.e. thus, they are $T_e$-dependent as seen in Table I). The last additive terms in Eqs.16 account for the energy density transfer from the NTH distribution to the TH electrons ($\frac{\partial U_{ee}}{\partial t}$) and lattice ($\frac{\partial U_{eL}}{\partial t}$), respectively. Therefore, thermalisation dynamics of the NTH electrons is incorporated into the model in the "source terms" to describe: (i) the generation of the thermalized electron subsystem and (ii) the interaction of NTH electrons with lattice. The two source terms include the features of the excitation pulse but they also incorporate the relaxation of the NTH electron population through Eq.8 [29].

### III. NUMERICAL SOLUTION

Due to the inherent complexity of Eqs.(1-16), an analytical solution is not feasible and therefore, a numerical approach is pursued. Numerical simulations have been performed using the finite difference method while the discretization of time and space has been chosen to satisfy the Neumann stability criterion. Furthermore, it is assumed that on the boundaries, von Neumann boundary conditions are satisfied and heat losses at the front and back surfaces of the material are negligible. The initial conditions are $T_e(t=0)=T_L(t=0)=300K$, while stresses, strains, and displacements are set to zero at $t=0$. Furthermore, the vertical stress $\sigma_{zz}$ of the upper surface is taken to be zero at all times (i.e. stress free). The parameters for Ni used in the simulation are summarised in Table I. The values of the laser beam features used in the simulation are: The (peak) fluence is $E_p \left(\equiv \sqrt{\pi}\tau_p I_0/\left(2\sqrt{ln2}\right)\right)$, where $I_0$ stands for the peak value of the intensity, spot radius $R_0$ (where the intensity falls to $1/e^2$) is equal to 15μm, and pulse duration values lie in the range [10fs, 2.3ps]. The wavelength of the beam is $\lambda_L$=800nm. We note that, for the sake of simplicity, the laser beam conditions considered in the first part of this work are selected so that: (i) they emphasise the role of the nonthermal electrons and (ii) they are not sufficient to induce material melting or plastic deformation. Hence, only elastic deformation of a portion of the material is assumed. Thus, three different values of the fluence, $E_p$=20mJ/cm², 40mJ/cm², and 60mJ/cm², were chosen to simulate the mechanisms that follow the above requirements. In principle, a common approach followed to solve problems that involve elastic displacements [50] or hydrodynamics [10] is the employment of a staggered grid finite difference method which is found to be effective in suppressing numerical oscillations. Unlike the conventional finite difference method, temperatures ($T_e$ and $T_L$) and normal stresses $\sigma_{ii}$ are computed at the centre of each element while time derivatives of the displacements and first-order spatial derivative terms are evaluated at locations midway between consecutive grid points. Furthermore, shear stresses $\sigma_{ij}$ are evaluated at the grid points. Numerical integration is allowed to move to the next time step provided that all variables at every element satisfy a predefined convergence tolerance of ±0.1%.

TABLE I: Parameters for Ni

| Parameter | Value |
|---|---|
| $C_e$ [$10^5$ J/m³K] | Fitting [37] |
| $C_L$ [J Kgr⁻¹ K⁻¹] | Fitting [51] |
| $G_{eL}$ [$10^{17}$ Wm⁻³K⁻¹] | Fitting [37] |
| $A$ [$10^7$ s⁻¹ K⁻²] | 0.59 [47, 48] |
| $B$ [$10^{11}$ s⁻¹ K⁻¹] | 1.4 [47, 48] |
| $k_{e0}$ [Jm⁻¹s⁻¹K⁻¹] | 90 |
| $T_{melt}$ [K] | 1728 |
| $T_0$ [K] | 300 |
| $\alpha'$ [$10^{-6}$K⁻¹] | 13.4 |
| $E$ [GPa] | 200 |
| $\nu$ | 0.31 |
| $\rho_L$ [Kgr/m³] | 8908 |
| $\varepsilon_F$ [eV] | 9.7 |

Furthermore, to explore damage threshold determination, a thermal criterion is applied in which Eqs.1-16 were used (by ignoring Eq.15 and elasticity contribution) for $E_p$=80mJ/cm² and 120mJ/cm². We note, that in a previous work, melting and disintegration of Nickel films were explored after irradiation with ultrashort pulses by using a combined atomistic-continuum modelling [39]. Certainly, a revised model that incorporates rTTM and atomistic modelling could enable a more precise damage threshold estimation, however, it is outside the scope of the present study.

## IV. RESULTS-DISCUSSION

The contribution of the nonthermal electrons and their interactions with the thermal electrons needs to be evaluated and be tested whether it is reflected on the optical property changes of the material and thereby the energy deposition and absorption. As a result, it is necessary to explore possible temporal changes of the reflectivity and absorption coefficient during the pulse duration before they are integrated into the rate equations (Eqs.16). In principle, a thorough investigation of the optical property changes is required as any change is expected to influence, firstly, excitation and relaxation processes, and secondly, the thermal and mechanical response of the material.

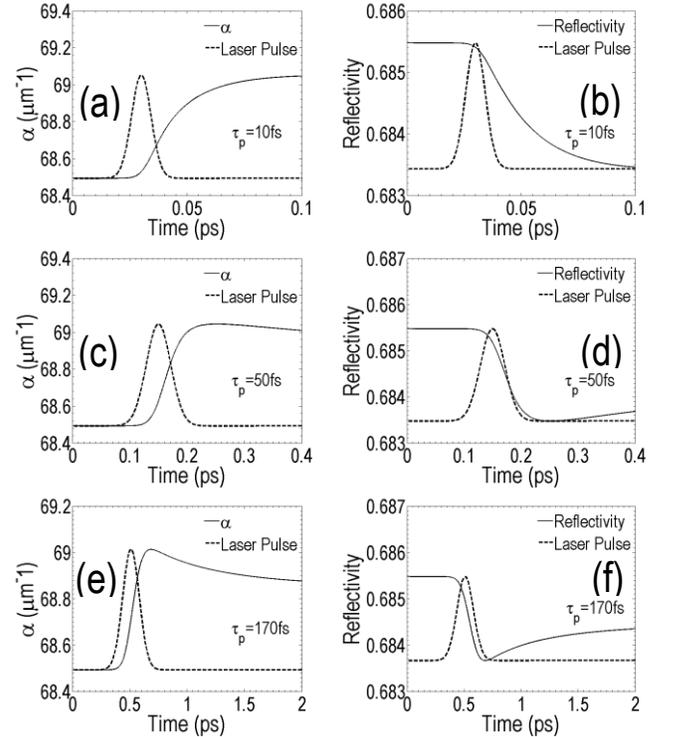

FIG. 3. (Color online) Temporal dependence of absorption coefficient and reflectivity for two different values of pulse duration, ($\tau_p$=10fs, 50fs and 170fs). ($E_p$ =40mJ/cm², 800nm laser wavelength, $R_0$=15μm).

Previous investigation of the transient optical properties after irradiation of noble metals (i.e. Au [46] or Cu [45]) demonstrates that there is a notable variation of both the reflectivity and the absorption coefficient that are also validated through pump-probe experiments performed by measuring reflectivity changes [41]. Similarly, our simulations for various pulse durations ($\tau_p$=10fs, 50fs and 170fs) using the rTTM on copper [52] also confirm a distinct variation of the magnitude during the pulse duration which is more pronounced at small pulse durations; such a behaviour implies that the absorbed energy changes within the pulse duration due to the significant reflectivity changes. Similarly, the large deviation of the absorption coefficient is expected to substantially affect the decay length of the electromagnetic wave and the laser energy localization. Therefore, it is evident that consideration of the optical properties variation is anticipated to lead to discrepancies in the thermal response of the system and the estimation of damage thresholds [45, 46]. Nevertheless, theoretical results for three different pulse duration values ($\tau_p$=10fs, 50fs and 170fs) indicate that the transient optical property changes after irradiation of Ni are very small during the period in which the laser is on (Fig.3) and therefore both reflectivity and absorption coefficient variations are not expected to influence substantially the thermomechanical response of the system. Similarly small changes are predicted for other values of the pulse duration or fluences ($E_p$=20mJ/cm$^2$ and 60mJ/cm$^2$) [52]. Nevertheless, to provide a more rigorous exploration of the fundamental processes and provide a material-based complete description, the calculated transient behaviour is incorporated into Eqs.16. Pump-probe experiments demonstrate an indicative large reflectivity change for copper in comparison to results for Ni which are in agreement with our simulations [41]. Similar small variations have also been observed in a more recent work [53].

Next, to determine the role of the nonthermal electrons in the heat exchange and relaxation process, it is important to provide a computed estimate of the energy density per unit time stored in the nonthermal electronic distribution. Fig.4 illustrates the power density which in the traditional TTM is stored in the electron system while the net result of the contribution of $\partial U_{ee}/\partial t$ and $\partial U_{eL}/\partial t$ indicates the large amount of overestimation of the internal energy of the thermal electrons (Fig.4a-d). Simulation results illustrated in Fig.4 indicate that compared to the TTM, the revised model predicts a delay of the heating process of both the TH electrons and the lattice. More specifically, the maximum power density transferred to the thermal electrons and lattice systems (*dashed* and *solid* lines, respectively) is shifted compared to whether it is assumed that instantaneous electron thermalisation occurs (*dashed dotted* line). It is evident from the temporal profile of the heat sources that the electronic dynamics is dominated by electron-electron scattering dominates in the initial stages.

On the other hand, compared to noble metals [29, 35], there exists a gradually significant nonthermal electron-lattice interaction for Ni which is attributed to the large electron-phonon coupling.

In contrast to pulse duration variation that is expected to influence the aforementioned power densities, variation of fluence is not expected to yield different values for the ratios between these quantities and therefore the shapes of the predicted temporal evolution are the same as those in Fig.4(a-d). This is attributed to the fact that fluence enters the power densities as a multiplicative coefficient and thereby it is cancelled out when relevant ratios are computed.

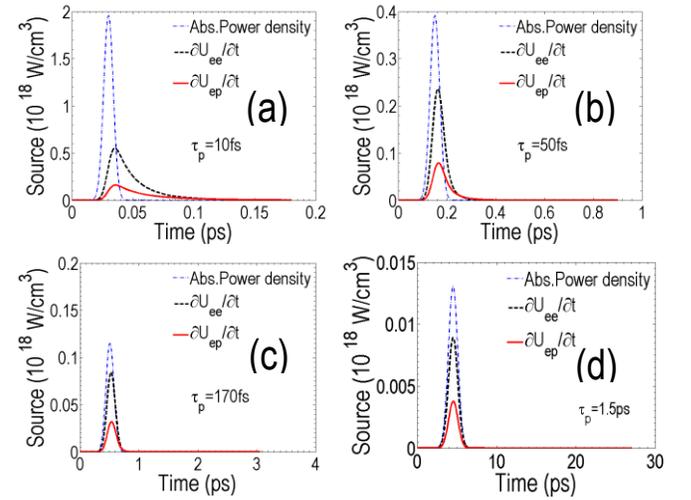

FIG. 4. (Color online) Time evolution of: (i) the laser source power density (*dashed-dotted* line), (ii) $\partial U_{ee}/\partial t$ (*dashed* line) and (iii) $\partial U_{eL}/\partial t$ (*solid* line) at maximum intensity for Ni irradiated with 10fs laser beams ($E_p$=40mJ/cm$^2$, 800nm laser wavelength, $R_0$=15μm). Simulations were performed for four values of $\tau_p$ (10fs (a), 50fs (b), 170fs (c), 1.5ps (d)).

Solution of Eqs.16 allows the investigation of the thermal response of the system through the analysis of thermalisation process and quantification of the evolution of $T_e$ and $T_L$. A comparison of the maximum surface electron and lattice temperatures as a function of time (for four different pulse durations, $t_p$=10fs, 50fs, 170fs, and 1.5ps) for $E_p$=40mJ/cm$^2$ simulated with the traditional TTM and rTTM is presented in Fig.5. It is evident by comparing the two models that the electron temperature peak is reached with a short delay with respect to the value attained if the TTM model is employed. This consequence is ascribed to the erroneously assumption of an instantaneous creation of thermal electrons and heating process of the electron and lattice baths predicted from the TTM model; in practice, there is a finite time required for the creation of the nonthermal electrons which leads to a delayed heating of both the thermal electrons and lattice (Fig.6). This argument is also supported by the temporal evolution of the power densities (Fig.4) and it is evident that delay is more pronounced for small pulse durations rather than longer excitation periods where the delay disappears ($\tau_p$=1.5ps). A similar behaviour is exhibited at different values of the fluence value ($E_p$=20mJ/cm$^2$ and 60mJ/cm$^2$) [52]. The employment of the revised model indicates that the incorporation of the contribution of the nonthermal electrons lowers the electron temperature. The decrease of the electron temperature is a physical outcome of the electron energy loss due mainly to the scattering of the nonthermal electrons from the electronic bath. By contrast, a comparison of the $T_e$ and $T_L$ evolution curves demonstrates that despite the delayed heating of the electron and lattice systems, respectively, electron-phonon relaxation is expected earlier if the rTTM is used (Fig.5). On the other hand, the electron-phonon interaction accounts for the increase of the lattice temperature, which is further enhanced by the term that describes the interaction of the nonthermal electrons with the lattice (i.e. $\partial U_{eL}/\partial t$).

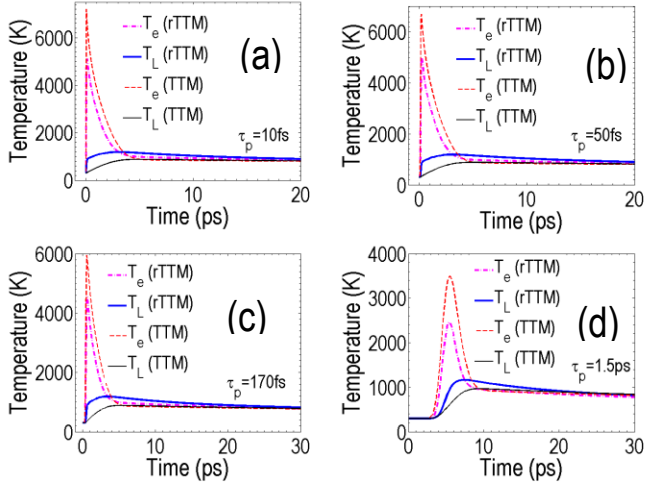

FIG. 5. (Color online) Electron and lattice evolution for $\tau_p$=10fs (a), 50fs (b), 170fs (c), and 1.5ps (d) derived from rTTM and TTM. ($E_p$ =40mJ/cm$^2$, 800nm laser wavelength, $R_0$=15μm).

In Fig.7, the TH electron population internal energy density $U_{TH}$ is also computed by means of the expression $C_e = n_e \partial \langle U_{TH} \rangle / \partial T_e$, ($n_e$ is the free electron density while $C_e$ is computed from Ref. [37]) if TTM or rTTM are used (*solid* line or *dashed* line) as a function of the pulse duration [54]. It is evident from the discrepancy of the internal energies that the energy stored in the NTH electron population is larger if the classical TTM is employed that underlines the overestimation due to the traditional approach. While TTM predicts an instantaneous thermalisation of the electrons, rTTM shows that the energy of the TH electrons (and therefore their population) is smaller due to the formation of a NTH population that exchanges energies with both the TH electrons and the lattice. Therefore, at later times (Figs.5-7) the TH electron population relaxes to the one predicted by TTM.

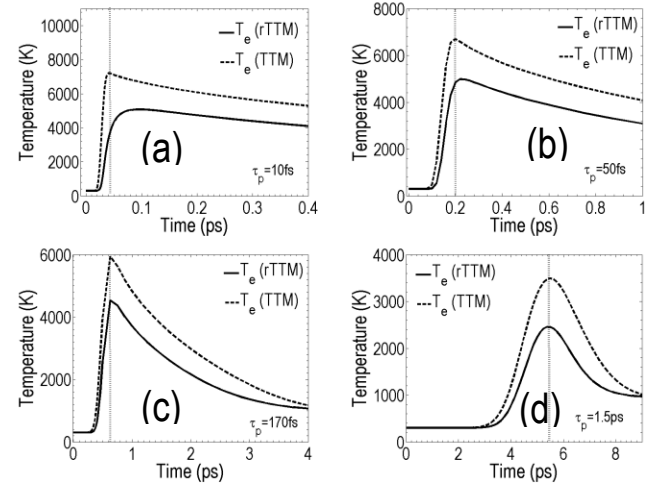

FIG. 6. Maximum electron temperature evolution computed using rTTM (*solid* line) or TTM (*dashed* line) for $\tau_p$=10fs (a), 50fs (b), 170fs (c), 1.5ps (d). Vertical *dotted* line indicates the heating delay of the electron system. ($E_p$ =40mJ/cm$^2$, 800nm laser wavelength, $R_0$=15μm).

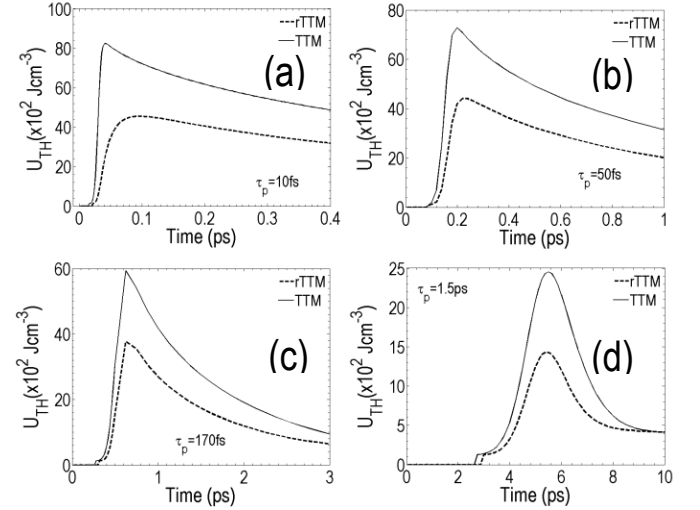

FIG. 7. Internal energy density stored in the TH electron population computed from TTM (*solid* line) and rTTM (*dashed* line) as a function of pulse duration ($E_p$ =40mJ/cm$^2$, 800nm laser wavelength, $R_0$=15μm).

Thermal response of the system has been quantified for various pulse duration values in the range [10fs, 2.3ps] for $E_p$=20mJ/cm$^2$, 40mJ/cm$^2$, 60mJ/cm$^2$, and the predicted values of the maximum electron and lattice temperatures are illustrated in Fig.8. The decrease of the maximum electron temperature with increasing pulse duration is illustrated in the graph while a small increase of the lattice temperature is produced. It is noted that the maximum lattice temperature discrepancy between the values predicted by TTM and rTTM increases for larger fluences. It is evident, though, that the computed maximum lattice temperature is below $T_{melt}$ that shows surface damage does not occur for the laser beam conditions of the simulations.

Variation of the lattice temperature induced by the absorption of the ultrashort optical pulse leads to mechanical effects due to the generation of thermal stresses. On the other hand, thermal stress produces strain generation and propagation. The spatio-temporal strain/stress pulse shape is determined by the solution of Eqs.15-16. To emphasise on the differences of the magnitude and spatial distribution of the strains and stresses predicted by the rTTM and TTM, the components of these fields along the direction of energy propagation ($z$-axis) are calculated and illustrated in Fig.9 and Fig.10, respectively. The strain $\varepsilon_{zz}$ is always positive at $z=0$ due to the fact the stress-free boundary condition does not imply any stretching perpendicular to the free surface. Furthermore, the solution of the second order (with respect to time) differential equation for the displacements (first equation of Eq.15) yields a two term propagating part, one showing a positive strain and a second with symmetric negative strain (result of the reflection on the surface). The two terms exhibit an exponential decay with length equal to $1/\alpha$. The exponential decay is reflected also on the stress fields which offer a more accurate calculation of the exponential decay length (Fig.10). Similar results for the strain pulse propagation have been presented in previous works for picosecond light pulses [55, 56]. Furthermore, reflectivity changes due to strain generation after ultrashort-pulsed laser irradiation of thin films on silicon surfaces have been also recently investigated [38, 57]. Comparing the strain values at $z=0$ with the results predicted in previous works ([55, 56]), the non-constant $\varepsilon_{zz}(z=0,t)$ at all times is attributed to the fact that in the current simulations, lattice temperature rise after irradiation neither occurs instantaneously nor remains constant.

It is evident by estimating the spatial position of the lowest strain or stress values at different times (Fig.9,10) that the strain and stress pulses propagate at a speed equal to approximately, 5578m/sec which corresponds to the longitudinal sound velocity in Ni ($=\sqrt{(2\mu+\lambda)/\rho_L}$ [56]). The pulses are illustrated at four different timepoints, $t=8$ps, 15ps, 20ps, and 30ps, after the arrival of the laser pulse on the surface of the material. In addition to $E_p=40$mJ/cm$^2$, simulations have been performed for $E_p=20$mJ/cm$^2$ and $E_p=60$mJ/cm$^2$ [52] where similar results (from a qualitative point of view) are deduced.

The comparison of the $\varepsilon_{zz}$ and $\sigma_{zz}$ pulses using rTTM and TTM demonstrates firstly the temporal shift of the waves that result from the delay of the lattice heating predicted by the revised model. Furthermore, the shapes of the strain and stress pulses derived from the two models do not appear to be significantly different at large timepoints.

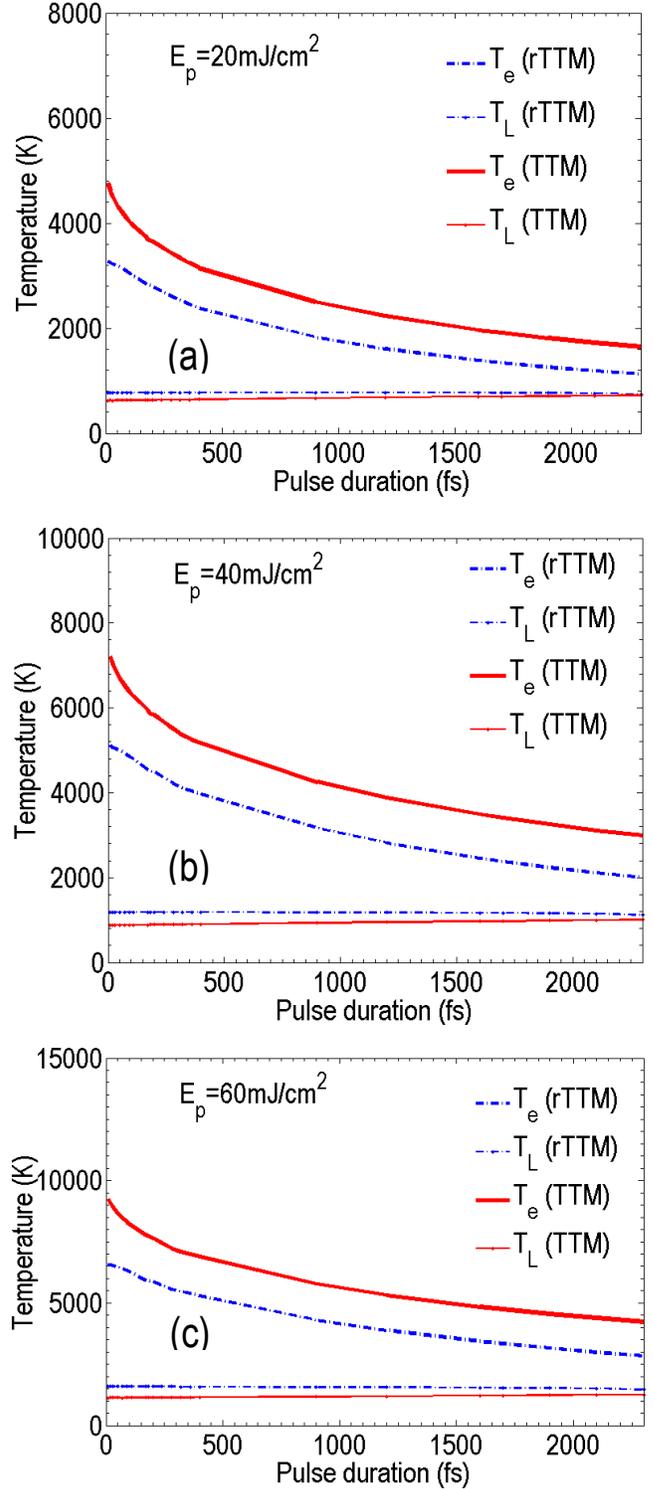

FIG. 8. (Color online) Maximum electron and lattice temperatures as a function of the laser pulse duration for three different fluences ($E_p$ =20mJ/cm$^2$,40mJ/cm$^2$,60mJ/cm$^2$, 800nm laser wavelength, $R_0$=15μm).

By contrast, there exists a larger than 20% increase in both the strain and stress size in a region *near* the surface at small times after laser irradiation (see Fig.9 and Fig.10 and videos in Supplementary Material [52]). A similarly substantially large deviation is also evident at bigger depths and more specifically, in the range [$5t/\alpha\sqrt{(2\mu+\lambda)/\rho_L}/120$, $5t/\alpha\sqrt{(2\mu+\lambda)/\rho_L}/80$]). The deviations are attributed to the enhanced lattice temperature values produced by the nonthermal electron interaction with lattice which is very pronounced at small timepoints while relaxation to values comparable to the ones predicted by the TTM is expected at bigger times (Fig.5).

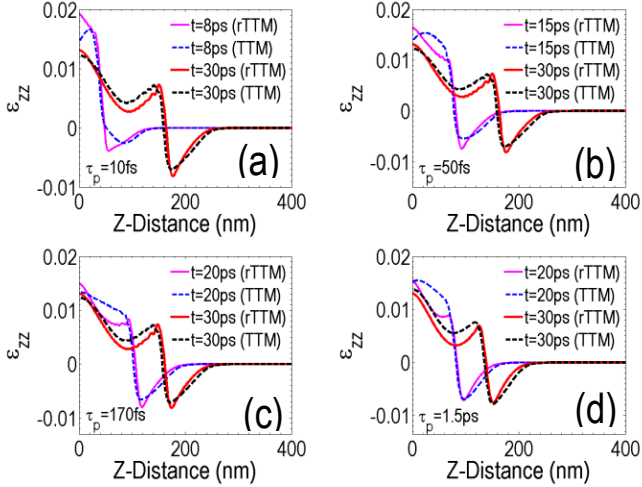

FIG. 9. (Color online) Spatial dependence of strain along *z*-direction (at *x*=0) at *t*=8ps, 15ps, 20ps, and 30ps for $\tau_p$=10fs, 50fs, 170fs, and 1.5ps derived from rTTM and TTM. ($E_p$ =40mJ/cm$^2$, 800nm laser wavelength, $R_0$=15μm).

An experimental validation of the proposed mechanism is required to test the adequacy of the theoretical model, however, the scope of this work is primarily related to the introduction of a consistent theoretical framework that will take into account: (i) the energy balance between the thermal electron and the lattice baths enriched with the contribution of nonthermal electrons, (ii) the influence of the DOS around the Fermi energy (Ni is characterised by a nonconstant DOS around $E_F$), (iii) a potential variation of the optical parameters during the pulse duration, (iv) the thermomechanical response of the material due to material heating. Nevertheless, theoretical investigation of the thermomechanical response of the material and comparison of the simulation results with experimental observables in previous works where a simpler version of the model was used [38, 57] confirm the primary importance and validation of the proposed underlying physical mechanism of laser matter interaction and associated processes. On the other hand, by ignoring the mechanical response of the system, previous works that assume a revised TTM based on the inclusion of nonthermal electron interaction with both thermalized electrons and lattice and the introduction of correction terms in the source term (as presented also in Ref. [29]) show a good agreement with results in pump-probe experiments [32-36] in noble metals.

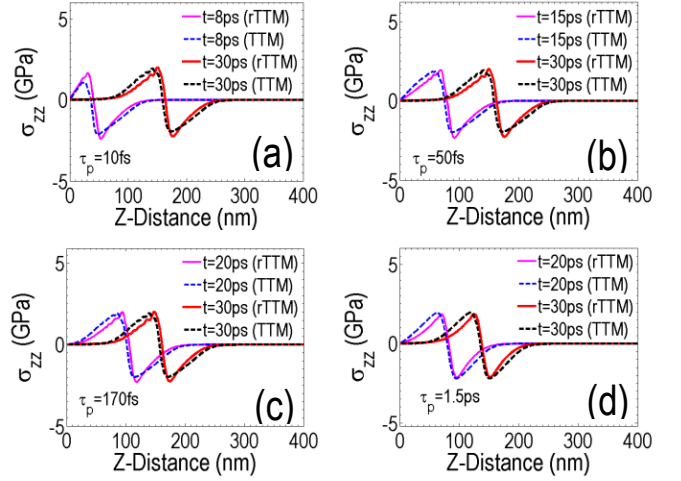

FIG. 10. (Color online) Spatial dependence of stress along *z*-direction (at *x*=0) at *t*=8ps, 15ps, 20ps, and 30ps for $\tau_p$=10fs, 50fs, 170fs, and 1.5ps derived from rTTM and TTM. ($E_p$ =40mJ/cm$^2$, 800nm laser wavelength, $R_0$=15μm).

Although, single shot laser irradiation at low fluence does not appear to influence the strain/stress fields shape (at larger times) if the role of the nonthermal electrons is taken into account, the picture, however, is expected to alter drastically in different laser conditions. For example, in multiple shot experiments of small temporal delays between the subsequent pulses (i.e. train-pulse technology), the approximately 20% variation of the strain wave amplitude or the substantially smaller electron temperature produced due to excitation could influence (through accumulation effects) surface micromaching techniques and applications. Furthermore, the significant deviation of the magnitude of the strain fields at small timepoints (i.e. ~8ps) could also influence mechanical properties of bilayered materials (for example, thin films on silicon surfaces [38, 57]) where strong acoustic waves are expected to be reflected on the interface and interfere strongly with the propagating strain leading to a more complex total strain.

The aforementioned simulations aimed, firstly, to emphasise the differences between theoretical predictions TTM and rTTM in conditions that do not induce phase transition. It is important to emphasise on the significant impact of the discrepancy between the predicted maximum lattice temperatures derived from the two models. This prospect is expected to have a very important impact on material properties and industrial applicability in terms of capability to modulate laser parameters as it will provide a more accurate and precise range of fluences to avoid surface damage. Regarding the employment of the model to explore mechanisms related to onset of damage, simulations have also been performed for fluences that lead to material melting. More specifically, the application of the rTTM model indicates that for $E_p$=80mJ/cm$^2$ or 120mJ/cm$^2$, the maximum lattice temperature exceeds $T_{melt}$ which suggests that material melts and surface modification occurs (Fig.11 and Fig.12). Maximum electron and lattice temperature dependence on pulse duration is the same as at lower fluences (Fig.8), however, the temperatures reached are higher and the discrepancy predicted by TTM and rTTM increases at larger laser energies (Fig.11). A comparison of the computed maximum lattice temperatures for all values of $E_p$ demonstrates that the maximum lattice temperature predicted using the revised model (rTTM) is always higher than that computed by means of TTM (Fig.12). It is evident that deviation of the maximum $T_L$ predicted from the revised model and the classical TTM increases as the fluence increases. This outcome is expected as at higher energies, a larger population of nonthermal electrons is produced and their role is enhanced. An estimation of the fluence that leads to temperatures larger than $T_{melt}$ derived from rTTM and TTM are 67mJ/cm$^2$ and 99mJ/cm$^2$, respectively. The 33% difference in the damage threshold determination is significant and it reveals the important thermal effects to the material due to the direct interaction of the nonthermal (hot) electrons with the lattice. Certainly, an experimental confirmation of the theoretical findings is necessary to validate the damage thresholds predicted by the model; nevertheless, the predicted substantial decrease of the damage threshold with respect to the estimation by means of the conventional TTM emphasises on the significant role physical processes in the early stages of irradiation which should not be neglected.

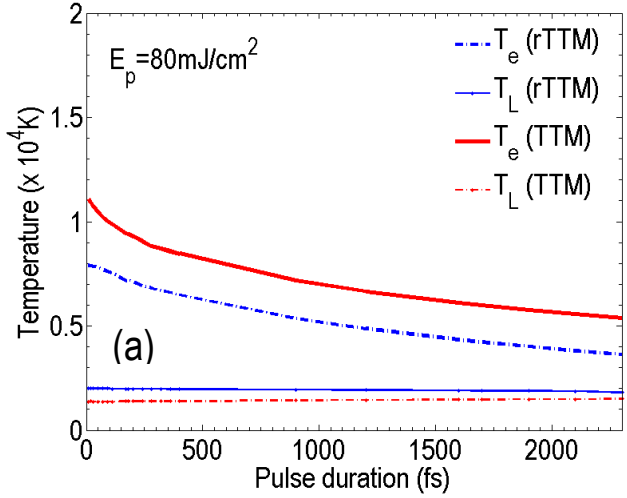

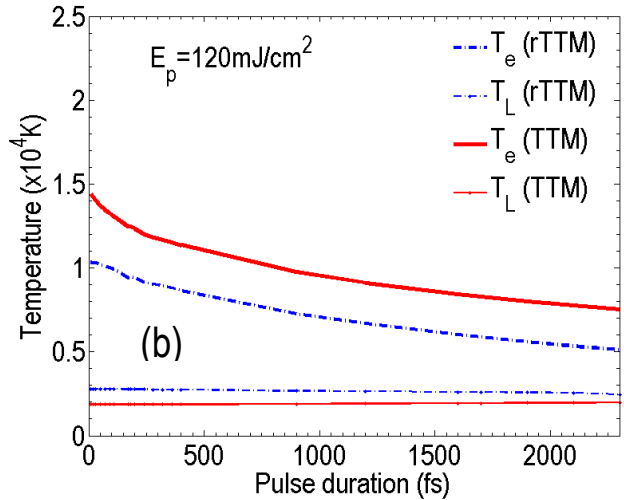

FIG. 11. (Color online) Maximum electron and lattice temperatures as a function of the laser pulse duration for two different fluences ($E_p$ =40 mJ/cm$^2$ and 120mJ/cm$^2$, 800nm laser wavelength, $R_0$=15μm).

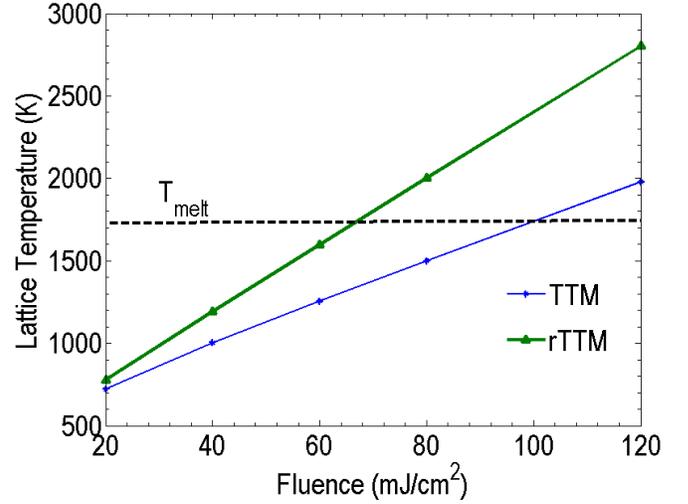

FIG.12. (Color online) Maximum lattice temperatures as a function of the laser pulse fluences (800nm laser wavelength, $R_0$=15μm) simulated with TTM and rTTM. *Dashed* line indicates the melting temperature $T_{melt}$.

## V. CONCLUSIONS

A detailed theoretical framework was presented that describes both the ultrafast dynamics of electrons and the induced strains and stresses in metals with strong electron-phonon coupling and complex DOS after irradiation with ultrashort pulsed lasers. The revised TTM incorporates the interaction of the nonthermal electrons with both the thermal electrons and the lattice while an additional component is included to describe the thermomechanical effects. A parametric analysis was performed for a range of pulse duration and fluence values that shows the differences from the predictions of the classical TTM. It is demonstrated that the employment of the revised TTM yields remarkably large values (~20%) for the induced strains especially at small timepoints after irradiation which is expected to be of paramount importance depending on the laser processing techniques (i.e. laser-pulse train processing). On the other hand, simulation results indicate also that the proposed underlying physical mechanism leads to a substantially (~33%) lower damage threshold for the irradiated material. This a very useful aspect from an industrial point of view towards estimating a more accurate damage threshold that is very significant for laser manufacturing approaches.

## ACKNOWLEDGEMENTS

G.D.T acknowledges financial support from *LiNaBioFluid* (funded by EU's H2020 framework programme for research and innovation under Grant agreement No 665337) and *Nanoscience Foundries and Fine Analysis (NFFA)–Europe* H2020-INFRAIA-2014-2015 (under Grant agreement No 654360).

*Corresponding author: tsibidis@iesl.forth.gr